\documentclass[]{spie}  
\usepackage[utf8]{inputenc}
\usepackage[english]{babel}
\usepackage{graphicx}
\usepackage{fancyhdr}
\usepackage{amsfonts}
\usepackage{appendix}
\usepackage{authblk}
\usepackage{url}
\usepackage[T1]{fontenc}
\usepackage{mathrsfs}  
\usepackage{amsmath,amssymb,stmaryrd}
\title{Ferroelectricity in tetragonal ZrO$_2$ thin films}

\author[1*]{Ali El Boutaybi}
\author[1]{Thomas Maroutian}
\author[1]{Ludovic Largeau}
\author[1]{Nathaniel Findling}
\author[2]{Jean-Blaise Brubach}
\author[2]{Rebecca Cervasio}
\author[1]{Alban Degezelle}
\author[1]{Sylvia Matzen}
\author[1]{Laurent Vivien}
\author[2]{Pascale Roy}
\author[3]{Panagiotis Karamanis}
\author[3]{Michel Rérat}
\author[1]{and Philippe Lecoeur}
\affil[1]{Centre de Nanosciences et de Nanotechnologies (C2N), Universite Paris-Saclay, CNRS, 91120 Palaiseau, France.}
\affil[2]{Synchrotron SOLEIL - CNRS - CEA Paris-Saclay, 91120 Palaiseau, France.}
\affil[3]{Université de Pau et des Pays de l'Adour, CNRS, IPREM, E2S UPPA, Pau, France.}
\affil[*]{ali.el-boutaybi@c2n.upsaclay.fr}

\begin{document} 
\maketitle
\begin{abstract}
We report on the crystal structure and ferroelectric properties of epitaxial ZrO$_2$ films ranging from 7 to 42 nm thickness grown on La$_{0.67}$Sr$_{0.33}$MnO$_3$-buffered (110)-oriented SrTiO$_3$ substrate. By employing X-ray diffraction, we confirm a tetragonal phase at all investigated thicknesses, with slight in-plane strain due to the substrate in the thinnest films. Further confirmation of the tetragonal phase was obtained through Infrared absorption spectroscopy with synchrotron light, performed on ZrO$_2$ membrane transferred onto a high resistive Silicon substrate. Up to a thickness of 31 nm, the ZrO$_2$ epitaxial films exhibit ferroelectric behavior, at variance with the antiferroelectric behavior reported previously for the tetragonal phase in polycrystalline films. However, the ferroelectricity is found here to diminish with increasing film thickness, with a polarization of about 13 $\mu$C.cm$^{-2}$ and down to 1 $\mu$C.cm$^{-2}$ for 7 nm and 31 nm-thick ZrO$_2$ films, respectively. This highlights the role of thickness reduction, substrate strain, and surface effects in promoting polarization in the tetragonal ZrO$_2$ thin films. These findings provide new insights into the ferroelectric properties and structure of ZrO$_2$ thin films, and open up new directions to investigate the origin of ferroelectricity in ZrO$_2$ and to optimize this material for future applications.
\end{abstract}
\rmfamily 
\section{INTRODUCTION} \label{s:intro} 
The discovery of ferroelectricity in Si-doped HfO$_2$ has expanded the research on HfO$_2$ and ZrO$_2$ beyond their conventional role as high-k dielectric materials \cite{Boscke2011}. These materials have been extensively investigated for various microelectronic applications \cite{Bauer2008} and are thermodynamically stable when in contact with silicon \cite{hubbard1996}, making them compatible with metal-oxide-semiconductor (CMOS) technology \cite{Copel2000, Müller2013}. Thanks to their ferroelectric behavior HfO$_2$ and ZrO$_2$-based compounds have demonstrated potential in various technological fields, offering opportunities for advancements in electronic devices \cite{Müller2013, Roy2020, Ali2018-FeFET, Park-energystorage2014, Hoffman-negative-c, PARK2019, Cheng2019, Wei2019, Cheema2022}. Antiferroelectricity has also been reported in ZrO$_2$ and ZrO$_2$-rich compounds, which further expands the prospects for these materials \cite{Muller2012, Park2015}.

Commonly reported polar phases of HfO$_2$ and ZrO$_2$ are the orthorhombic $Pbc2_1$ (\textit{o}-phase) and rhombohedral $R3m$ (\textit{r}-phase) phases \cite{Boscke2011, Wei2018}. However, these polar phases are metastable, and several approaches have been employed to stabilize these ferroelectric phases. One commonly used approach is through doping and controlling the grain size \cite{Starschich2017, Lee2019}. Doping alone may not necessarily stabilize the polar phase \cite{Batra2017, Dutta2020}, but it can alter its surface energy, which in turn may affect its stability and sometimes enhance the polarization. Therefore, even with doping, the grain size still plays a role in stabilizing the ferroelectric state in these materials, with surface effects playing a crucial role \cite{Park2017-NANO, Materlik2015, Ali2021}.

In recent years, the ferroelectric properties of pure ZrO$_2$ have garnered considerable attention, primarily due to its resemblance to HfO$_2$, as well as its greater abundance in nature compared to HfO$_2$ \cite{Lenzi2022}. Notably, investigations of ZrO$_2$ have been conducted in various forms, such as pure ZrO$_2$ in polycrystalline thin films, leading to the identification of the \textit{o}-phase \cite{Shibayama_2020, HUANG2021, Xu2022, Crema2023}. The same \textit{o}-phase has also been observed in epitaxial thin films \cite{Song2021}, for which both the \textit{r}-phase \cite{Silva2021, Ali2021} and the tetragonal $P4_2/nmc$ phase (\textit{t}-phase) \cite{Mimura2021} have been reported as well. Indeed, when the grain size is reduced to the nanoscale (< 30 nm), the stable phase of ZrO$_2$ is the non-polar \textit{t}-phase \cite{Garvie1965, Garvie1978, Pitcher2005}. This is attributed to the lower surface enthalpy of \textit{t}-ZrO$_2$ compared to the ground state monoclinic P2$_1$/c phase (\textit{m}-phase) of ZrO$_2$ \cite{Pitcher2005}. However, in thinner ZrO$_2$ films, polar phases (\textit{o} and \textit{r}) have been reported, as mentioned above. Thus, the stabilization of polar phases in these nanocrystalline materials involves a competition between various polymorphs, including both polar and non-polar phases. It is worth noting that increasing the thickness without maintaining and controlling a proper grain size can lead to degradation and suppression of the ferroelectric behavior \cite{Riedel2016, Zheng2022}.

In addition to the ferroelectric state, antiferroelectricity has also raised significant interest recently, particularly in ZrO$_2$-rich thin films. However, the origin of antiferroelectric phases is less studied compared to ferroelectric ones. First, orthorhombic $Pbca$ could potentially contribute to the observed antiferroelectric behavior in ZrO$_2$ \cite{Osamu1990}, as this phase is a combination of two unit cells of the polar \textit{o}-phase with opposite polarizations \cite{Materlik2015, Hyun2020, Kersch2021}. This phase is also found to be energetically close to the ground state \textit{m}-phase \cite{Kersch2021, Azevedo2022}. Additionally, the \textit{t}-phase has been identified as a potential driver of antiferroelectric behavior in ZrO$_2$ polycrystalline thin films \cite{Reyes2014}; this has been explained as a transition from the non-polar \textit{t}-phase to the polar \textit{o}-phase \cite{Reyes2014}. This second approach is the most commonly adopted in the literature \cite{Park2015, Wang2018, Xu2022, Dahlberg2022, Lomenzo2023}.

In this study, we demonstrate the ferroelectricity of \textit{t}-phase ZrO$_2$ thin films with varying thicknesses ranging from 7 to 42 nm. First, the \textit{t}-phase of ZrO$_2$ is evidenced by X-ray Diffraction (XRD). Then, synchrotron Infrared (IR) absorbance spectroscopy confirms the same \textit{t}-phase. To better understand the IR bands in our experimental results, we used first-principles calculations with the CRYSTAL suite of programs to investigate the IR activities and simulate the IR absorbance of \textit{t}-ZrO$_2$ \cite{CRYSTAL17}. The details about the calculations can be found in Ref. \cite{ElBoutaybi-JMC-C}. Finally, we show that ZrO$_2$ thin films exhibit ferroelectric behavior up to a thickness of 31 nm, indicating the development of polarization in a non-polar \textit{t}-phase when the film thickness is reduced. In line with studies on \textit{r}- and \textit{o}-phase epitaxial thin films \cite{Wei2018, Song2021, Silva2021, Ali2021}, the ferroelectric behavior is here obtained in the pristine \textit{t}-phase films, directly after growth, and thus do not correspond to a change to another crystalline phase upon field cycling \cite{Mimura2021}.
\section{EXPERIMENTAL SECTION} \label{s:2}
{\noindent}\textbf{Sample Fabrication.} Thin films of pure ZrO$_2$ ranging from 7 to 42 nm thickness were grown on (110)-oriented SrTiO$_3$ (STO) substrate with a La$_{0.67}$Sr$_{0.33}$MnO$_3$ (LSMO) buffer layer using pulsed laser deposition (PLD). The deposition was performed at a temperature of 800 °C, under an O$_2$ growth pressure of 0.1 mbar for ZrO$_2$ and 0.18 mbar for LSMO films, and with a laser energy fluence of 1.1 J.cm$^2$ at a repetition rate of 2 Hz. The target-substrate distance was maintained at 6 cm. The as-grown films were cooled down to room temperature at a rate of 10 °C/min.\\
Due to the large IR response of STO \cite{Maletic_2014}, the IR measurements were performed on a ZrO$_2$ membrane, which was obtained by chemically etching the LSMO bottom electrode to release the ZrO$_2$ film from the STO substrate. Subsequently, the ZrO$_2$ film was transferred onto a high-resistivity silicon substrate (Si-HR) for IR measurement.

{\noindent}\textbf{Structure analysis.} The crystal structure and epitaxial orientation of the thin films were analyzed using XRD techniques. For $\theta-2 \theta$ out-of-plane scans, a Panalytical X’pert Pro diffractometer was utilized (Figure \ref{fig:1}), while other XRD analyses were performed using a Rigaku SmartLab diffractometer. Both diffractometers were configured in a parallel beam setup and employed monochromated Cu $K \alpha_1$ radiation with a wavelength of 1.54059 Å.

In addition, IR spectroscopy was employed to investigate the phonon activities of the ZrO$_2$ membrane, specifically the IR absorbance of ZrO$_2$ in transmission configuration. The IR measurements were performed at the AILES beamline of synchrotron SOLEIL with a Bruker IFS125 FT-IR spectrometer. Spectra in the far infrared region ( 20 cm$^{-1}$ to 700 cm$^{-1}$ ) were obtained by using a 6 µm-thick Mylar beam splitter and a 4.2$K$ bolometer detector.  Absorbance spectra were recorded with a resolution of 2 cm$^{-1}$, at 4 cm.s$^{-1}$ mirror speed, and by averaging 200 scans \cite{ROY2006139}.

{\noindent}\textbf{Ferroelectric characterization.} To measure the ferroelectric properties of the films, platinum/ZrO$_2$(film)/LSMO capacitors were fabricated using standard photolithography and lift-off techniques. The capacitors were then characterized using a ferroelectric tester (AixACCT, TF analyzer 1000). The film's ferroelectric response was obtained through Positive Up Negative Down (PUND) measurements. The polarization-electric field (P-E) loops were measured at a frequency of 1 kHz with a triangular waveform.
\section{RESULTS $\&$ DISCUSSION} \label{s:3}
\subsection{ZrO$_2$ thin films}
\begin{figure}[h]
    \centering
     \hspace*{-1.3in}
    \includegraphics [scale=0.7] {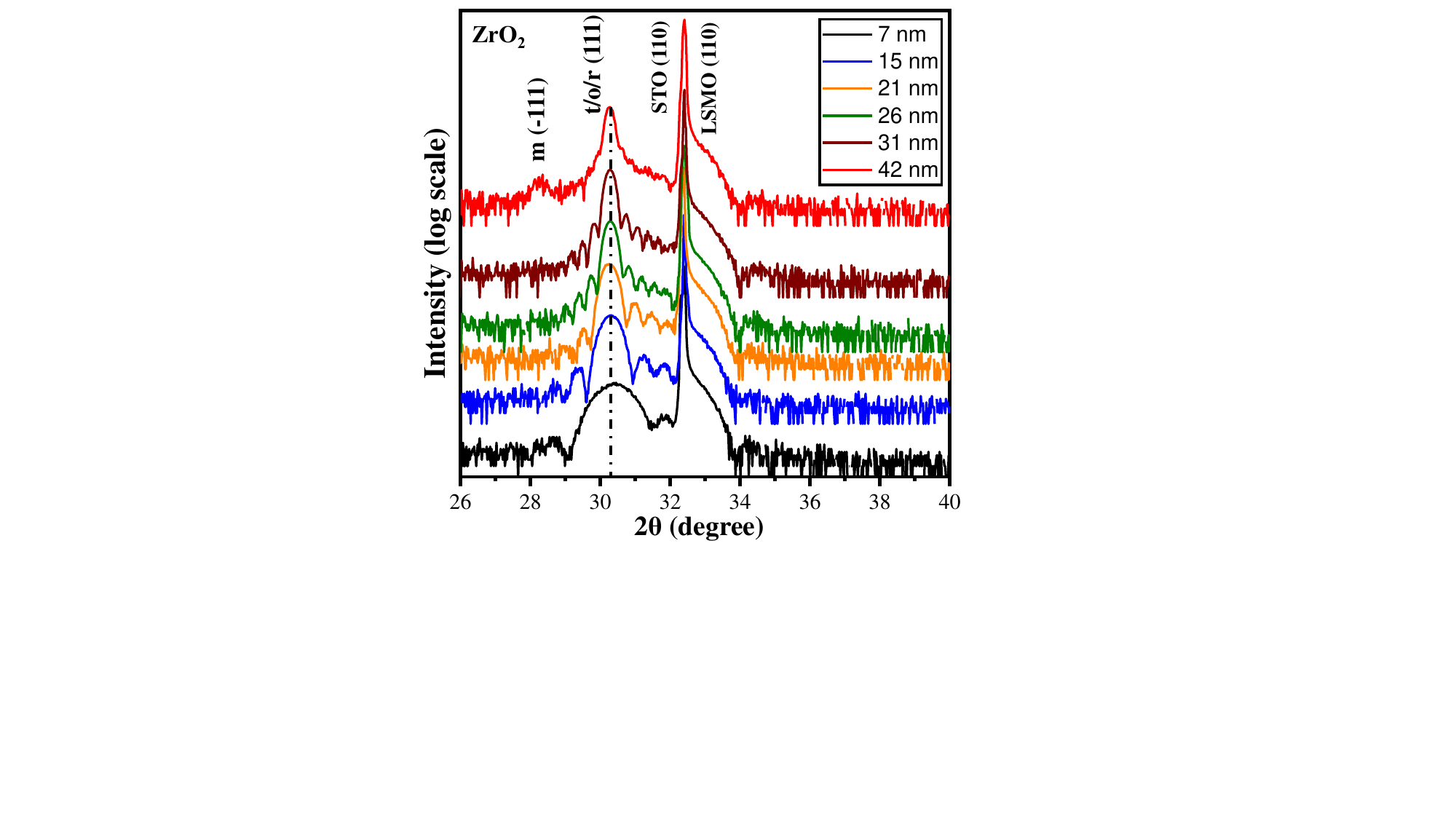}
    \vspace*{-47mm}
    \caption{Out-of-plane X-Ray diffraction, $\theta - 2 \theta$ of pure ZrO$_2$ at different thicknesses.}
    \label{fig:1} 
\end{figure}
Figure \ref{fig:1} shows the $\theta-2\theta$ out-of-plane XRD of ZrO$_2$ thin films at different thicknesses ranging from 7 to 42 nm. The most intense peaks correspond to the (110)-oriented STO substrate, while the peaks closest to the substrate correspond to the (110) pseudo-cubic orientation of the LSMO bottom electrode. The thickness of LSMO was determined to be close to 20 nm using X-ray reflectivity. At a 2$\theta$ angle of about 30.30°, the (111) peak of ZrO$_2$ is observed, which can correspond to the \textit{o}-, \textit{r}-, or \textit{t}-phases. It should be noted that this peak around 30° is ascribed to the \textit{o}-phase or \textit{r}-phase when the films exhibit ferroelectric behavior \cite{Boscke2011, Song2021,  Wei2018, Ali2021, Silva2021}. In Figure \ref{fig:1}, it can be observed that for a ZrO$_2$ film of approximately 42 nm thickness, the ground state \textit{m}-phase is observed at $2\theta$ = 28.32°; Similar thickness dependence has been reported for ZrO$_2$ \textit{r}-phase \cite{Ali2021}. In all ZrO$_2$ thin films, the (111) peak is observed at the same 2$\theta$ of about 30.30°. Only in the thinnest film of 7 nm, this peak appear slightly shifted to the right, indicating a small tensile strain in the film, as expected for the (110)-STO substrate \cite{Jiao2021}.

\begin{figure*}[h]
    \centering
     \hspace*{-1.2in}
    \includegraphics [scale=0.66] {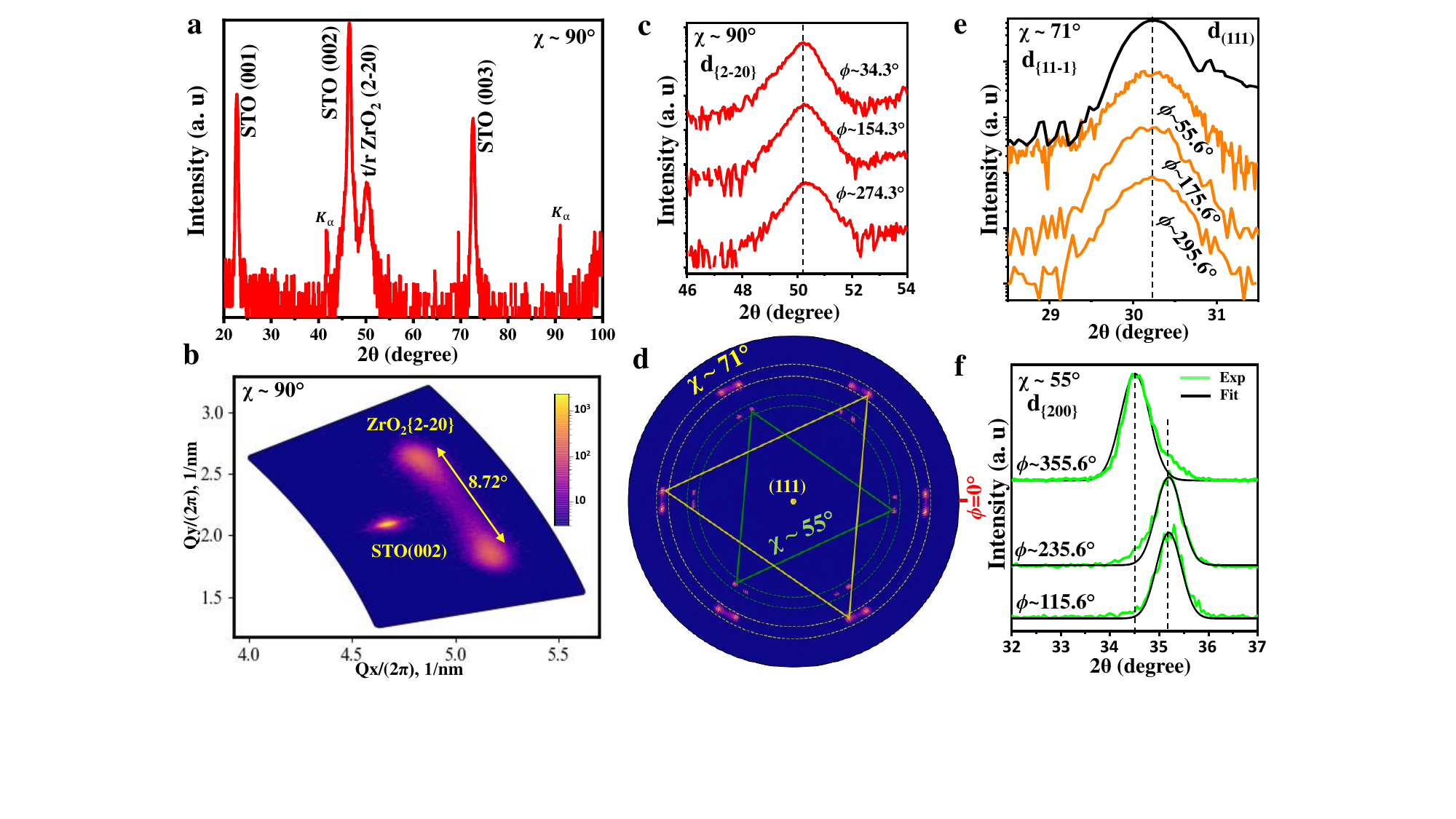}
    \vspace*{-25mm}
    \caption{X-Ray diffraction measurements performed on a 15 nm-thick pure ZrO$_2$ thin film: (a) in-plane $\theta - 2 \theta$ scan along the <001> STO azimuthal direction, (b) reciprocal space map ($2\theta - \phi$ scan) around the (002) STO plane, (c) in-plane $\theta - 2 \theta$ scan for one variant observed in (b), (d) pole figures measured at $2 \theta = 30.30^\circ$ and $34.5^\circ$, (e) $\theta - 2 \theta$ scan for one variant observed at around $\chi = 71^\circ$ (highlighted in yellow in pole figure), and (f) $\theta - 2 \theta$ scan for one variant observed at $\chi = 55^\circ$ (highlighted in green in pole figure).}
    \label{fig:2} 
\end{figure*}
Further XRD measurements were conducted to gain insight into the phase and symmetry of ZrO$_2$ thin films. In particular, to distinguish between the different possible phases of ZrO$_2$, specific families of crystallographic planes can be analyzed. For instance, the $\{1-10\}$ planes are not present in the \textit{t}-phase, whereas they are present in the \textit{m}-, \textit{o}-, and \textit{r}-phases. In the \textit{r}-phase, these planes are all present but with low intensity due to the high symmetry of the phase, and, accordingly, these planes were not reported in ZrO$_2$ \textit{r}-phase \cite{Ali2021}. However, the \textit{r}-phase has equal lattice parameters, a = b = c, which make it distinguishable from other phases. The same is true for the \textit{m}-phase, which can be easily identified from XRD analysis (Figure \ref{fig:1} in red). On the other hand, the \textit{o}-phase has a lattice parameter close to the one of the \textit{t}-phase, making it challenging to distinguish it from the \textit{t}-phase based only on the unit cell lattice parameters. Thus, the main difference by XRD between the  \textit{t}- and \textit{o}- phases is the presence of the $\{1-10\}$ family planes in the \textit{o}-phase \cite{Yun2022}.
First, to obtain the film orientation with respect to the substrate and detect any other phases, in-plane XRD measurements were conducted along the <001> of STO azimuthal direction. These measurements are known to be more sensitive to sample size than film thickness, allowing to detect other phases that may be present at the film/substrate interface or in small amounts in the thin film, as previously observed in ZrO$_2$ and Hf$_{0.5}$Zr$_{0.5}$O$_2$ (HZO) on (001)-oriented STO substrate \cite{Ali2021}. Figure \ref{fig:2}.a displays the in-plane XRD of 15 nm-thick ZrO$_2$ along <001> azimuthal direction, where an additional peak at a $2\theta$ of approximately 50.15° is observed. This peak can correspond to the \textit{r}- or \textit{t}-phase. For the \textit{o}-phase, a peak at 24.5° is expected for the (1-10) plane \cite{Yun2022}, although only one plane from the $\{1-10\}$ family of $Pbc2_1$ phase is allowed by symmetry reasons.

It is worth noting that the alignment of ZrO$_2$ films with the substrate does not follow a specific direction of the latter. Indeed, the peak observed at 50.15° is not precisely aligned with the [001] direction of the STO substrate. An in-plane 2D ($2\theta - \phi$) map was performed around the (002)-STO plane, as presented in Figure \ref{fig:2}.b, demonstrating that the (2-20) plane is rotated by about 4.3° relative to the [001]-STO direction. This finding is also supported by the pole figures generated at $2\theta =$ 34.5° and $2\theta =$ 30.30° (Figure \ref{fig:2}.d). The appearance of two spots in Figure \ref{fig:2}.b indicates the existence of another $\{2-20\}$ plane family, which is rotated by 8.72° relative to the first spot. This epitaxial relationship has also been reported in HZO films on (110)-STO substrates with a similar angle of rotation (8.5°) \cite{Song2022-110}. The split into different variants observed at a given angle (Figure \ref{fig:2}.b) may represent a way for the ZrO$_2$ film to minimize its energy. Additionally, it is important to mention that ZrO$_2$ thin films (and HfO$_2$ as well) do not exhibit direct epitaxy on perovskite substrates but rather follow a domain-matching epitaxy \cite{Estandia2020}. Further, the in-plane $\phi$ scan around $\{2-20\}$ planes results in 12 peaks (see Figure S1 in supplementary information). Taking one variant and performing Bragg scans (Figure \ref{fig:2}.c) reveals the close inter-atomic distance between these planes (d$_{(20-2)} \sim$ d$_{(0-22)} \sim$ 1.885 Å and d$_{(2-20)} \sim$ 1.871 Å), suggesting that the phase of ZrO$_2$ is tetragonal.

The absence of the $\{1-10\}$ family planes in Figure \ref{fig:2}.a indicates that the peak at $2\theta$ $\sim$ 50.15° corresponds to the \textit{t}-phase, although the possibility of the \textit{r}-phase cannot be entirely ruled out, given its low structure factor intensity for $\{1-10\}$ planes (see Table S1). To verify that the absence of $\{1-10\}$ planes is not solely due to in-plane measurements taken in a specific azimuthal direction (<001> of STO), an identical in-plane 2D ($2\theta - \phi$) map, as the one shown in Figure \ref{fig:2}.b, was performed around the (001)-STO, and the $\{1-10\}$ plane is expected to diffract at around 24.5° if present. The result is provided in the supplementary information (Figure S1), confirming the absence of the $\{1-10\}$ plane. Furthermore, to ensure that the $\{1-10\}$ plane is absent throughout the entire film, a pole figure was conducted around a $2\theta$ angle of 24.5° (Figure S1), which also confirms the lack of signal for the $\{1-10\}$ plane. Notably, the $(1-10)$ of the \textit{m}-phase was easily observed in thick ZrO$_2$ (42 nm) and 10 nm-thick pure HfO$_2$ (see supplementary information), exhibiting a structure factor intensity close to that of the \textit{o}-phase (see Table S1). Finally, it should be noted that the $(1-10)$ plane of the \textit{o}-phase was readily identified in 10 nm-thick Y-doped HfO$_2$ \cite{Yun2022}. Therefore, the absence of detection of the (1-10) plane in the 15 nm-thick ZrO$_2$ film supports the conclusion that its phase does not belong to the orthorhombic system.

At this stage, the close relationship between the \textit{t}- and \textit{r}- phases, particularly when the film is (111) oriented, underscores the necessity for additional structural analysis to comprehend the crystal structure of ZrO$_2$. Even though in the \textit{r}-phase, the d$_{(2-20)}$, d$_{(2-20)}$, and d$_{(2-20)}$ should be similar, the 2-fold symmetry of (110)-oriented STO substrate can induce such difference as observed in Figure \ref{fig:2}.c. For this, one can investigate the $\{11-1\}$ planes, which provide information about the angle of the unit cell \cite{Wei2018}. In the \textit{r}-phase, the difference in inter-atomic distance between d$_{(111)}$ and d$_{\{11-1\}}$ gives a unit cell angle below 90° \cite{Wei2018, Ali2021}. In contrast, in the \textit{t}- and also in the \textit{o}-phase, d$_{(111)}$ and d$_{\{11-1\}}$ are equal, yielding a unit cell angle of 90°. 

To investigate the $\{11-1\}$ planes, a pole figure was generated at $2\theta =$ 30.30°, and the result is shown in Figure \ref{fig:2}.d (12 spots are found at $\chi \sim$ 71°). The pole figure confirms the angle between the two twins (about 8.7°), as observed in the in-plane 2D maps (Figure \ref{fig:2}.b), and reveals the presence of four ZrO$_2$ variants in the film. Focusing on one variant, the Bragg scans ($\theta -2\theta$ scans) of the $\{11-1\}$ planes are provided in Figure \ref{fig:2}.e. The inter-atomic distances of the $\{111\}$ (out-of-plane) and the $\{11-1\}$ (at $\chi \sim$ 71°) planes are almost the same, with only a small tensile distortion, giving a unit cell angle of about 90.07° (+/- 0.05). This small distortion of the unit cell can be induced by tensile strain caused by the substrate. This measurement indicates that the phase is not rhombohedral. 

Based on previous investigations, the ZrO$_2$ film on (110)-STO is tetragonal. To determine the lattice parameters of this tetragonal unit cell, a pole figure was performed at $2\theta = 34.5^\circ$, and is given in Figure \ref{fig:2}.d (highlighted in green). One variant was selected to examine the $a$, $b$, and $c$ parameters, while the other variants are given in the supplementary information. Figure \ref{fig:2}.f displays the $\theta-2\theta$ scans for one variant, showing two similar peaks corresponding to $a$ and $b$ lattice parameters, and a peak at a clearly different $2\theta$ value corresponding to the $c$ parameter of the \textit{t}-phase. The extracted lattice parameters are $a\sim b=5.096 Å$ and $c=5.195 Å$, which fall close to the values reported for \textit{t}-ZrO$_2$ in the literature \cite{Kisi1998}. 

Furthermore, Figure \ref{fig:2}.f shows that the peak at $\phi=235.6^\circ$ appears to extend on the lower $2\theta$ side, while the peak at $\phi=355.6^\circ$ has a shoulder on the higher $2\theta$ side, suggesting the possibility of mixed lattice parameters in one crystallographic direction. In the 42 nm-thick ZrO$_2$ film (see Supplementary information), double peaks were indeed observed along each of the 3 azimuths, indicating the presence of two lattice parameters. This finding suggests the existence of not only 4 but 12 variants in 42 nm-thick ZrO$_2$. This was also observed in 26 nm-thick ZrO$_2$ film, as will be discussed below (Figure \ref{fig:3}). This increase in variant number with increasing film thickness suggests that this is a way for the ZrO$_2$ thin films to relax the strain. It is important to note that similar double peaks were also observed in a 10 nm-thick Y-doped HfO$_2$ thin film \cite{Yun2022}. Here, in the case of pure ZrO$_2$, they are hardly observed in a 15 nm-thick film (Figure \ref{fig:2}.f) and clearly observed in thicker (>15 nm) ones (see Figure \ref{fig:3} and supplementary information).

Table \ref{table:1} displays the extracted lattice parameters of ZrO$_2$ at different thicknesses. The lattice parameters are slightly higher at a thickness of 15 nm than at 42 nm, indicating a small tensile strain in thinner films. As the thickness increases to 42 nm, the lattice parameters approach those reported for ZrO$_2$ in powder form, as reported in Ref. \cite{Kisi1998} (also listed in Table \ref{table:1}).
\begin{table} [h]
    \centering
     \caption{Lattice parameters a, b, and c of 15, 26, and 42 nm-thick ZrO$_2$ thin films. *Lattice parameters of 26 nm-thick ZrO$_2$ membrane.}
\begin{tabular}{ c c c c c } 
\hline
materials&  a(Å)  & b(Å)  &  c(Å) \\ [1ex] 
\hline\hline
ZrO$_2$ (15 nm)&5.096&5.096& 5.195\\[0.5ex] 
ZrO$_2$ (26 nm)&5.088&5.089& 5.183\\[0.5ex] 
ZrO$_2$ (26 nm)*&5.089&5.089& 5.184\\[0.5ex] 
ZrO$_2$ (42 nm) &5.085&5.085&5.187\\
ZrO$_2$ \cite{Kisi1998} &5.074&5.074&5.188\\
\hline 
\hline 
\end{tabular}
 \label{table:1}
\end{table}
\subsection{ZrO$_2$ membrane}
\begin{figure*}[h]
    \centering
     \hspace*{-0.6in}
    \includegraphics [scale=0.56] {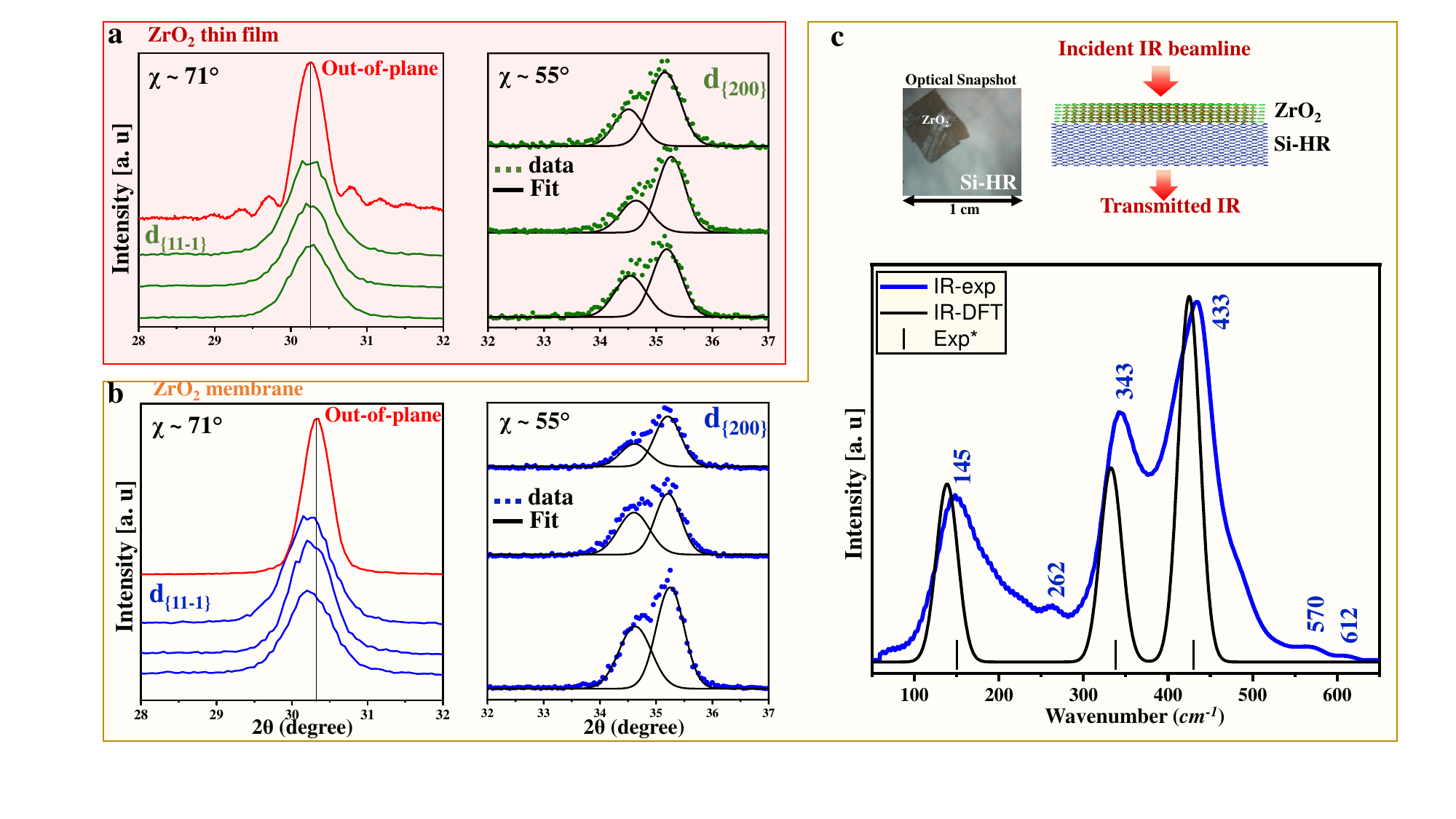}
    \vspace*{-11mm}
    \caption{X-Ray diffraction and IR measurements performed on a 26 nm-thick pure ZrO$_2$:  (a) $\theta - 2 \theta$ Bragg scans performed on ZrO$_2$ thin film for $\{111\}$  and $\{200\}$ crystal planes at $\chi = 71^\circ$ and $\chi = 55^\circ$, respectively. (b) Same Bragg scans obtained on ZrO$_2$ membrane for the $\{111\}$ and $\{200\}$ crystal planes at $\chi = 71^\circ$ and $\chi = 55^\circ$, respectively. (c) Experimental IR absorbance of ZrO$_2$ membrane, with the peak positions (in cm$^{-1}$) indicated in blue. The IR absorbance obtained using the CRYSTAL code is also provided for the \textit{t}-phase. *Additionally, earlier experimental IR results for the \textit{t}-phase taken from Ref. \cite{BONERA2003} are indicated.}
    \label{fig:3} 
\end{figure*}
In addition to XRD measurements, IR spectroscopy offers an alternative method for characterizing the crystal structure of materials by examining their phonon activities, which act as distinctive fingerprints for each crystal symmetry. Notably, IR spectroscopy has demonstrated its capability to identify and differentiate between different ZrO$_2$ phases \cite{Kersch2022, ElBoutaybi-JMC-C}. To perform IR characterization, a 26 nm-thick ZrO$_2$ film was transferred onto a Si-HR substrate (as shown in the snapshot insert in Figure \ref{fig:3}.c). The use of Si-HR, which is transparent in the IR region, allows access to the absorbance response of ZrO$_2$, facilitating precise analysis of its properties.

Before conducting IR absorbance measurements, XRD analyses were performed both before and after the transfer of the ZrO$_2$ film to investigate the impact of the release on the film structure. Figures \ref{fig:3}.a and b depict $\theta - 2 \theta$ Bragg scans carried out on the ZrO$_2$ thin film and ZrO$_2$ membrane, focusing on the $\{11-1\}$  and $\{200\}$ crystal planes at $\chi = 71^\circ$ and $\chi = 55^\circ$, respectively. As previously mentioned, the $\{200\}$ $\theta - 2 \theta$ Bragg scans reveal the presence of double peaks in thick ZrO$_2$, indicating a combination of two lattice parameters in one given crystallographic direction. The extracted lattice parameters are displayed in Table \ref{table:1}, confirming a tetragonal phase in both ZrO$_2$ film and membrane. However, a noteworthy distinction emerges when examining the $\theta - 2 \theta$ Bragg scans around $\{111\}$, indicating no distortion in the ZrO$_2$ thin film with unit cell angle $\sim$ 90° (+/- 0.05°), while in the ZrO$_2$ membrane, a distorted unit cell angle $\sim$ 90.20° (+/- 0.05°) is obtained, corresponding to tensile strain. Conversely, the lattice parameters $a$, $b$, and $c$ exhibit minimal changes upon releasing the ZrO$_2$ film to obtain the membrane, as illustrated in Table \ref{table:1}.

The utilization of XRD enabled us to confirm that releasing a ZrO$_2$ film from its substrate does not alter its crystal structure, but does induce some additional strain, as evidenced in Figures \ref{fig:3}.a and b. Subsequently, the \textit{t}-phase of ZrO$_2$ was further substantiated through IR absorbance spectroscopy. As presented in Figure \ref{fig:3}.c, it revealed the presence of the three main characteristic bands of the \textit{t}-phase of ZrO$_2$ \cite{Materlik2015, ElBoutaybi-JMC-C}, in excellent agreement with earlier experimental findings reported in Ref. \cite{BONERA2003} (also depicted in bar form in Figure \ref{fig:3}.c). The IR absorbance of the ZrO$_2$ \textit{t}-phase is also confirmed by our theoretical IR absorbance calculated using the CRYSTAL17 package, as given in black in Figure \ref{fig:3}.c. Specifically, the experimentally observed values of the three main IR bands, as indicated in Figure \ref{fig:3}.c, are in good agreement with the theoretical ones at 139, 333, and 426 cm$^{-1}$, for $E_u$ , $A{2u}$ , and $E_u$ IR phonon modes of the ZrO$_2$ \textit{t}-phase, respectively. The simulated theoretical intensities also closely match the experimental ones (Figure \ref{fig:3}.c). Furthermore, the two lower IR bands with low intensity also agree well with previous calculations, as reported in Refs. \cite{Materlik2015, ElBoutaybi-JMC-C}.

In addition to the clear observation of the three characteristic IR absorbance bands of the \textit{t}-phase, three additional bands are detected on the IR absorbance spectrum in Figure \ref{fig:3}.c, albeit with low intensities. These additional bands could be attributed to the strain induced during the releasing and transferring of the ZrO$_2$ film, as evident in the XRD results shown in Figure \ref{fig:3}.b. The transfer process primarily induced tensile strain in the film, influencing the unit cell angle and, to a very small extent, the lattice parameters of the \textit{t}-ZrO$_2$ (Table \ref{table:1}). The strain effect on the IR absorbance of ZrO$_2$ has been previously discussed, and it was demonstrated that strain activates and amplifies new IR bands for the \textit{t}-phase of ZrO$_2$ \cite{ElBoutaybi-JMC-C}. It should here be underlined, that the IR absorbance can clearly be used to distinguish between different phases of ZrO$_2$, such as \textit{r} , \textit{o} , \textit{t}, and \textit{m} phases \cite{Materlik2015, ElBoutaybi-JMC-C}. Based on the present results from XRD and IR absorbance, and in line with previous ones \cite{ElBoutaybi-JMC-C}, it is evident that ZrO$_2$ remains tetragonal under tensile strain, albeit with a small distortion observed in thinner films. 
\subsection{Ferroelectric measurements}
\begin{figure*}[h]
    \centering
     \hspace*{-0.8in}
    \includegraphics [scale=0.60] {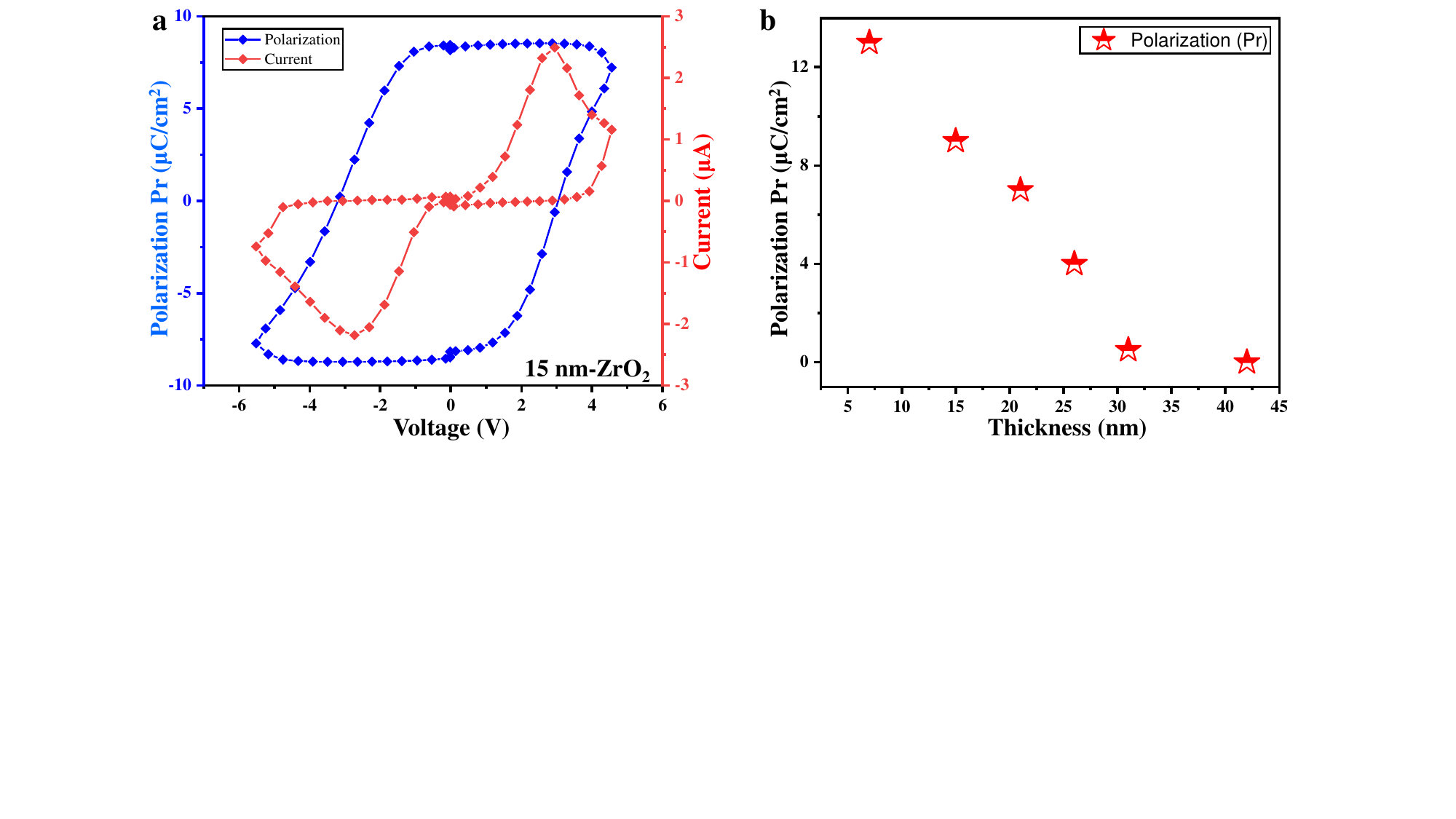}
    \vspace*{-56mm}
    \caption{(a) Polarization hysteresis loop of 15 nm-thick pure ZrO$_2$ thin film, and (b) remanent polarization Pr value at different thicknesses (see supplementary information for the corresponding polarization and switching current loops).}
    \label{fig:4} 
\end{figure*}
From our XRD and IR spectroscopy data (Figures \ref{fig:2} and \ref{fig:3}), the investigated epitaxial ZrO$_2$ thin films in this study are shown to be in the tetragonal phase. Electrically, these ZrO$_2$ films demonstrate ferroelectric properties up to a thickness of 31 nm. It is important to underline that, at variance with the study of Mimura \textit{et al.} \cite{Mimura2021}, no field cycling is here needed to obtain a ferroelectric signature in our \textit{t}-phase ZrO$_2$ epitaxial thin films. For instance, Figure \ref{fig:4}.a presents the P-E loops for the 15 nm-thick ZrO$_2$ thin film in its pristine growth state, exhibiting a clear ferroelectric behavior with a polarization of approximately 9 $\mu$C.cm$^{-2}$. Notably, this polarization value is lower than the one reported for pure \textit{r}-ZrO$_2$ films within the same thickness range (21 $\mu$C.cm$^{-2}$) \cite{Ali2021}. This discrepancy can be attributed to the presence of compressive strain in the \textit{r}-ZrO$_2$ phase \cite{Silva2021, Ali2021}, whereas the films in our study are under small tensile strain. Nevertheless, Figure \ref{fig:4}.a unambiguously demonstrates the clear ferroelectric nature of the 15 nm-thick ZrO$_2$ film. Similar P-E loops are obtained on all our films up to 31 nm thickness, and do not show any electric-field induced change upon repeated measurements. These findings underline the significance of strain in inducing the ferroelectric behavior in ZrO$_2$ thin films.

The ferroelectric polarization of the ZrO$_2$ thin films at various thicknesses was further investigated using the same measurement as shown in Figure \ref{fig:4}.a. The remanent polarization (Pr) at different thicknesses is reported in Figure \ref{fig:4}.b, indicating that Pr gradually decreases with increasing film thickness. Specifically, the 7 nm-thick ZrO$_2$ film exhibits a remanent polarization of approximately 13 $\mu$C.cm$^{-2}$, while the value reduces to around 1 $\mu$C.cm$^{-2}$ for the 31 nm thickness. For the 42 nm-thick ZrO$_2$ film, no ferroelectric behavior is observed, suggesting that the ZrO$_2$ \textit{t}-phase becomes fully relaxed, consequently suppressing any polarization effect induced by the in-plane strain.


\subsection{Discussion}
In the case of pure ZrO$_2$, ferroelectricity has been reported in the \textit{r}-phase \cite{Silva2021, Ali2021} and in the \textit{o}-phase, reports on the latter being the most numerous.  Experimental studies on polycrystalline thin films of pure ZrO$_2$ have shown polarizations in the range of 7-13 $\mu$C.cm$^{-2}$ \cite{Starschich2017, Shibayama_2020, Xu2022, Crema2023}, which is lower than the polarization theoretically predicted for the ZrO$_2$ \textit{o}-phase, giving a high value above 50 $\mu$C.cm$^{-2}$ \cite{Materlik2015, Ali-PRB-2023}. In epitaxial thin films of o-ZrO$_2$, a polarization of about 27 $\mu$C.cm$^{-2}$ has been reported \cite{Song2021}. To our knowledge, the only report of ferroelectricity in \textit{t}-phase ZrO$_2$ is from Mimura \textit{et al.} \cite{Mimura2021}; however, the initial pristine film exhibited an antiferroelectric response, and after field cycling, the film became ferroelectric with a XRD signature from the \textit{o}-phase \cite{Mimura2021}. In contrast, our films, as shown in Figure \ref{fig:4}, exhibit wake-up free behavior, indicating that they are inherently ferroelectric. This wake-up free behavior has consistently been observed in epitaxial thin films, regardless of their phases \cite{Wei2018, Song2021, Ali2021, Silva2021}. However, none of the polar phases mentioned earlier were observed in our ZrO$_2$ films; instead, only a tetragonal phase with tensile distortion was evidenced, suggesting that the induced ferroelectricity is primarily driven by substrate strain and thickness effects.

An  application of strain on the \textit{t}-phase of ZrO$_2$ will result in the disruption of its symmetry, and could consequently induce polarization. For example, a distorted \textit{t}-phase described as $Cm$ monoclinic polar phase was found theoretically, with a polarization close to the one of the \textit{r}-phase and a very similar IR response compared to the \textit{t}-phase \cite{ElBoutaybi-JMC-C}. Possibly, this polarization could be enhanced by strain as observed in the \textit{r}- and \textit{o}-phases \cite{Wei2018, Estandía2019, Silva2021}. Other effects can induce ferroelectricity in ZrO$_2$. In particular, Lenzi \textit{et al.} showed that oxygen vacancies can promote a polarization in the \textit{t}-phase of ZrO$_2$ \cite{Lenzi2022}. Oxygen migration was also found to contribute to the measured ferroelectricity at reduced thickness, as reported in \textit{r}-HZO and \textit{o}-ZrO$_2$ thin films \cite{Pavan2021, Li2023}.

Antiferroelectricity has also been observed in ZrO$_2$ thin films of the same \textit{t}-phase \cite{Park2015, Wang2018, Xu2022}, which contrasts with our findings in epitaxial films. This discrepancy indicates that the occurrence of antiferroelectricity in polycrystalline films may be influenced by microstructure or grain boundaries rather than being an intrinsic property of ZrO$_2$ \textit{t}-phase. Consequently, more investigations are necessary to elucidate the origin of polarization in the \textit{t}-phase of ZrO$_2$, and to understand the impact of epitaxial versus polycrystalline structures. More generally, further research is needed to uncover the complex interplay between microstructure, film growth, and the emergence of ferroelectricity or antiferroelectricity in ZrO$_2$ thin films.
\section {Conclusion}
We investigated the structure and ferroelectric behavior of epitaxial ZrO$_2$ thin films at different thicknesses grown on (110)-SrTiO$_3$ substrate. XRD results showed a \textit{t}-phase of ZrO$_2$, with a small distortion in thinner films. The \textit{t}-phase of ZrO$_2$ was further confirmed by its characteristic IR absorbance spectrum. The films ranging from 7 to 31 nm thickness exhibited a clear ferroelectric behavior, right in their pristine state after growth, with a decreasing remanent polarization with increasing film thickness. Our results thus evidence that a distorted non-polar \textit{t}-phase displays ferroelectric behavior at reduced thickness, and provide insights into the structure of ZrO$_2$ and its ferroelectric behavior. This discovery paves the way for exploring the origins of the ferroelectricity in ZrO$_2$-based compounds, and improve their performance for future applications.

\section*{Conflicts of interest}
There are no conflicts to declare.

\section*{Acknowledgements}
This work has received support from the Agence Nationale de la Recherche (ANR) under projects FOIST (N°ANR-18-CE24-0030) and FLEXO (N°ANR-21-CE09-0046), as well as from the French national network RENATECH for nanofabrication. Part of this work was granted access to the HPC resources of [CCRT/CINES/IDRIS] under the allocation 2021-2022 [AD010807031R1 ] made by GENCI (Grand Equipement National de Calcul Intensif). We also acknowledge the “Direction du Numérique” of the “Université de Pau et des Pays de l’Adour” and the mésocentre Aquitain (MCIA) for their computing facilities.\\
\\
\bibliographystyle{spiebib}
\bibliography{references}
\newcommand{\beginsupplement}{%
    \setcounter{table}{0}
    \renewcommand{\thetable}{S\arabic{table}}%
    \setcounter{figure}{0}
    \renewcommand{\thefigure}{S\arabic{figure}}%
    \renewcommand{\thesection}{S\arabic{section}}
}
\appendix 
\beginsupplement 
\section{Supplementary Information}
\begin{figure}[H]
    \centering
     \hspace*{-1.0in}
    \includegraphics [scale=0.65] {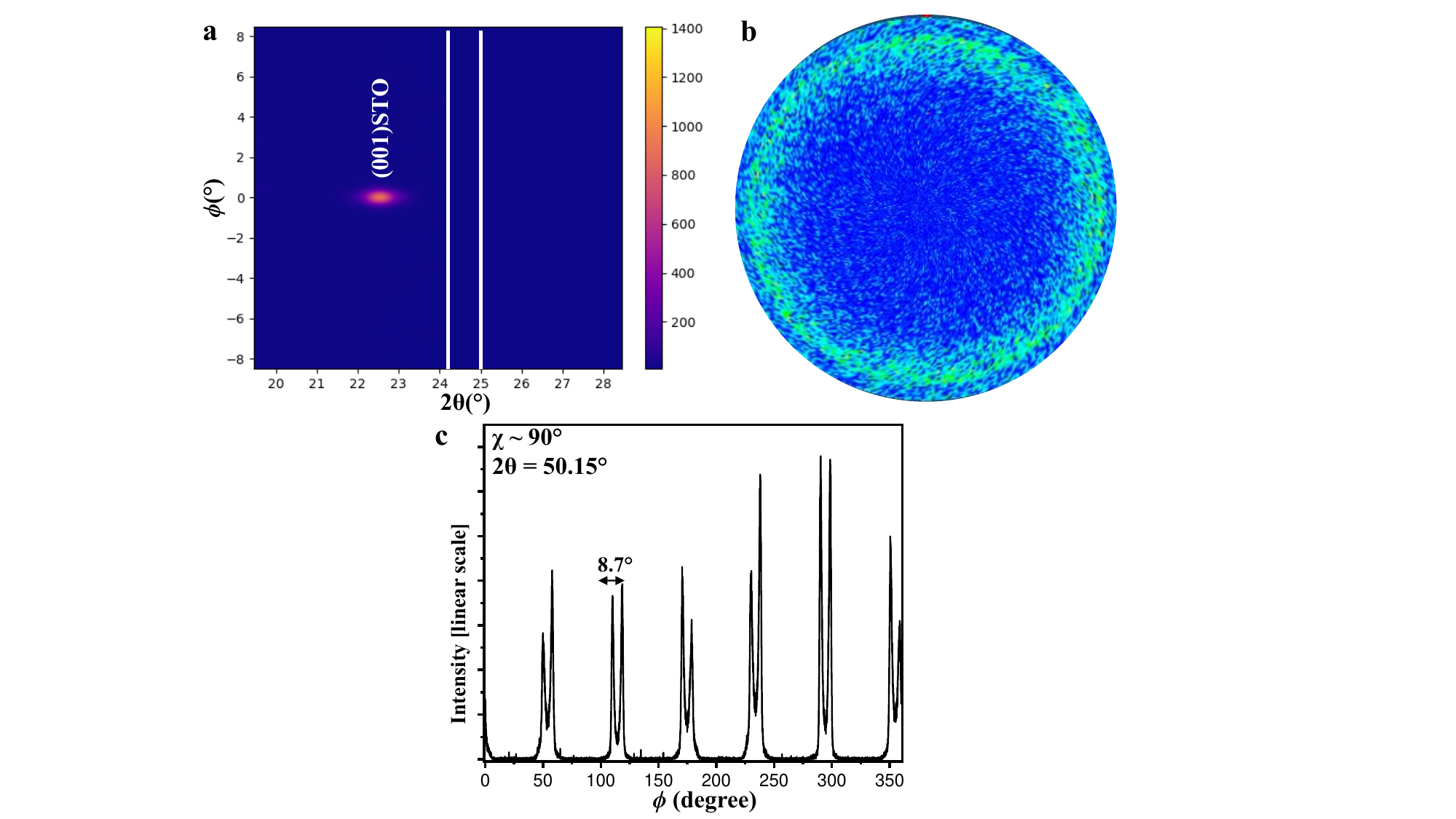}
    \vspace*{-5mm}
    \caption{(a) In-plane pole figure generated at  $2 \theta$ = 24.5°. (b) In-plane $\phi$ scan performed around  $2 \theta$ = 50.15° (see the main paper). }
    \label{fig:S1} 
\end{figure}

\noindent \textbf{Structure factors.}\\
Table \ref{table:S1} presents the structure factors of different ZrO$_2$ polymorphs obtained using Python xrayutilities package. Both the m and o phases present enough intensity to be detected by XRD measurement, but only the $\{1-10\}$ planes of m phase are observed (Figures \ref{fig:S4} and \ref{fig:S7}).
\begin{table} [H]
    \centering
     \caption{Comparison of the structure factors of m, t, o, and r$-$ phases for selected crystallographic planes.}
\begin{tabular}{ c c c c c c} 
\hline
&  m  & t  &  o  &  r & c\\ [1ex] 
\hline\hline
I$_{(111)}$ &109&204&196& 206&  116\\[0.5ex] 
I$_{(2-20)}$&84& 201&158&  170  &  108 \\[0.5ex] 
I$_{(1-10)}$&33& 0&45&  6  &   0 \\[0.5ex] 
I$_{(1-10)}$/I$_{(2-20)}$&0.39&0&0.28&0.03&0\\ 
  \hline 
\end{tabular}
 \label{table:S1}
\end{table}
\noindent \textbf{15 nm thick-ZrO$_{2}$.}\\
Figure \ref{fig:S1}.a presents an in-plane RSM (Reciprocal Space Map) 2D map ($2\theta - \phi$), similar to the one discussed in the main paper, but taken around the (001) plane of STO. The expected $\{1-10\}$ planes, if the o phase was present in the film, should be around 24.5° as indicated by the two lines in Figure \ref{fig:S1}.a, and they are not detected here.\\
\\
Furthermore, Figure \ref{fig:S1} shows a pole figure performed at $2\theta$ = 24.5° to search for the o phase. Once again, no $\{1-10\}$ planes were detected, even though they should appear at around $\chi$ = 35° if the o phase was present in the film.

\begin{figure}[H]
    \centering
     \hspace*{-0.9in}
    \includegraphics [scale=0.6] {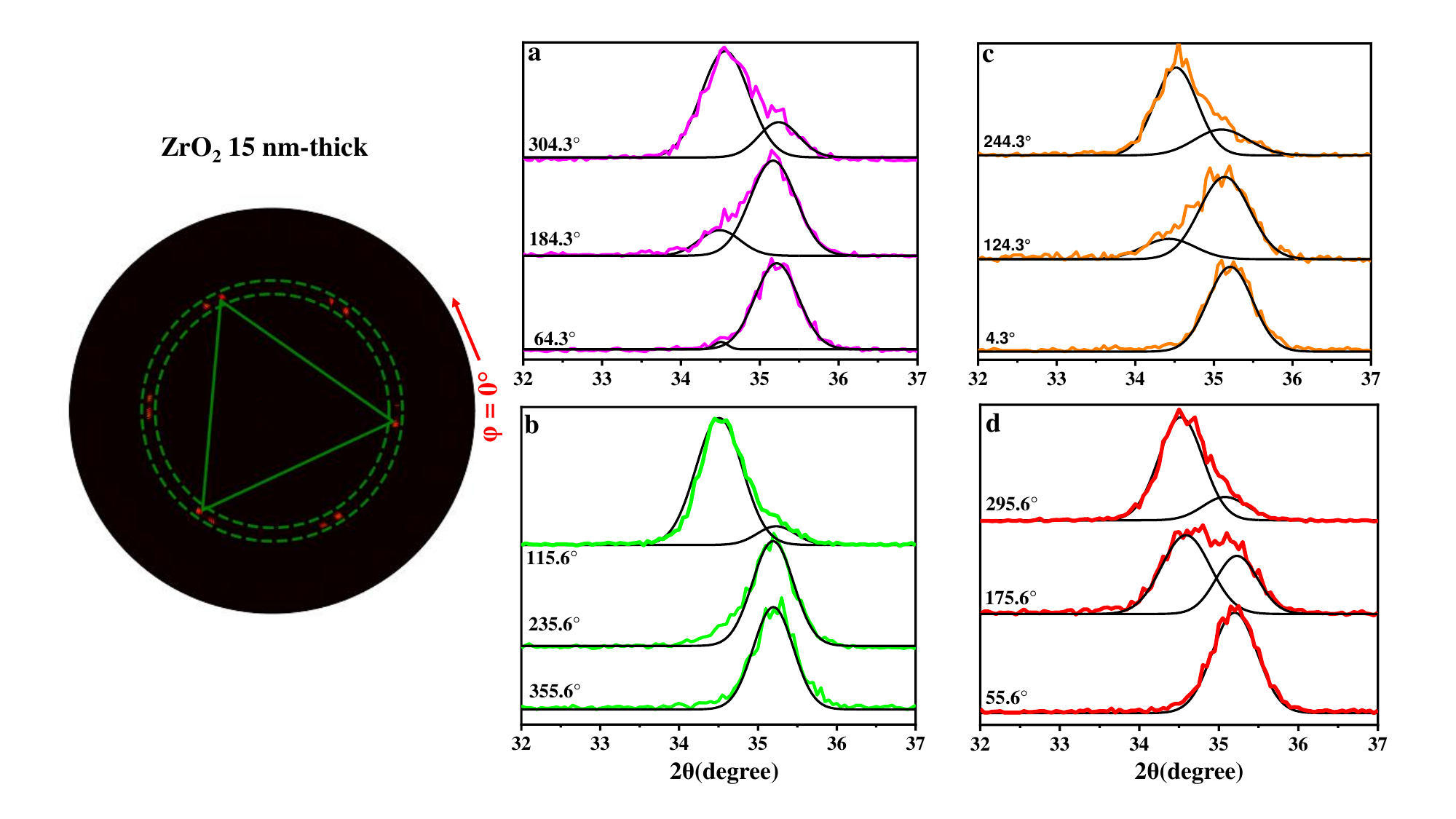}
    \vspace*{-8mm}
    \caption{X-ray diffraction pole figure generated at $2\theta=34.5^\circ$, showing 12 spots at around $\chi$=55°. The Bragg $\theta - 2\theta$ scan for each spot observed in the pole figure is shown on the right.}
    \label{fig:S2} 
\end{figure}

\begin{figure}[H]
    \centering
     \hspace*{-0.5in}
    \includegraphics [scale=0.45] {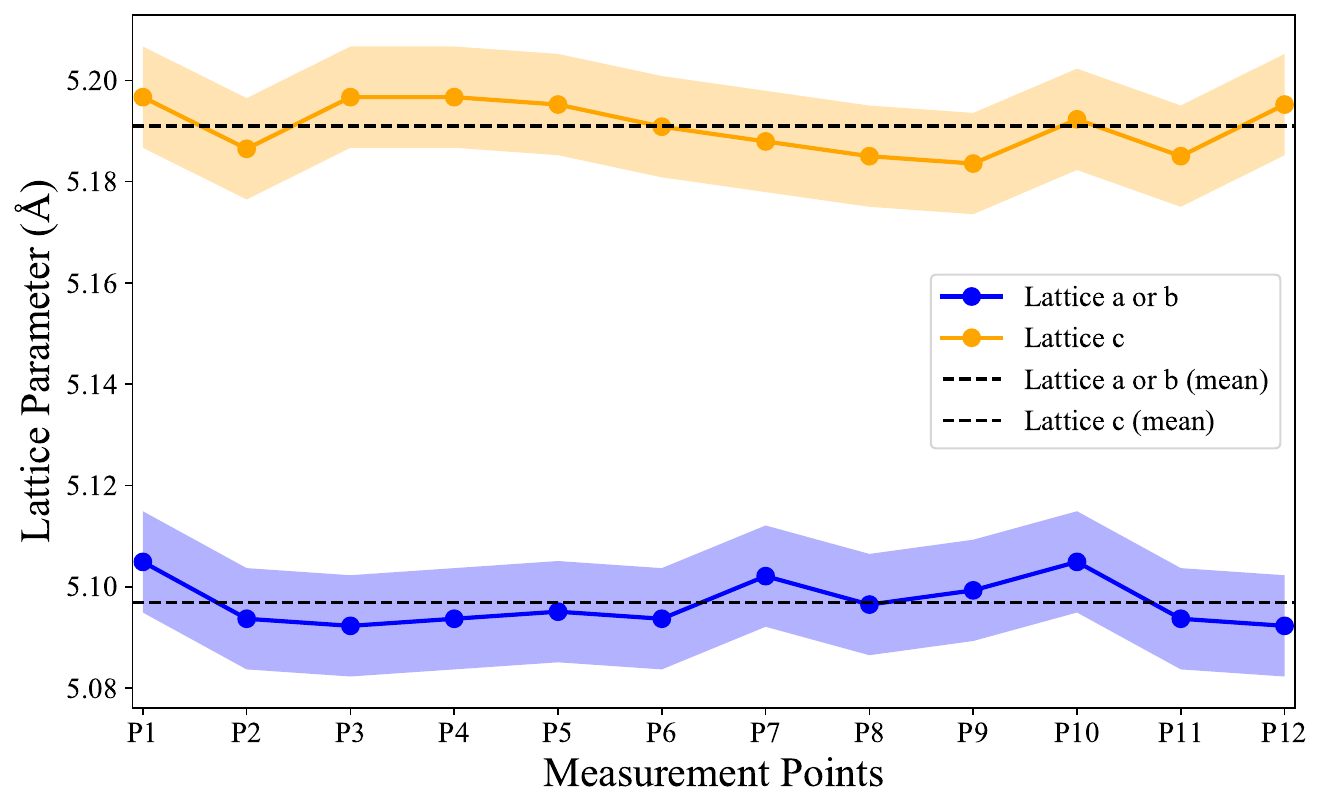}
    \vspace*{-2mm}
    \caption{The extracted a, b, and c parameters from Figure \ref{fig:S2}, with P1 to P12 corresponding to spots at $\phi =$ 3.8° (first spot after $\phi =$ 0°) to $\phi =$ 355°, respectively.}
    \label{fig:S3} 
\end{figure}

\begin{figure}[H]
    \centering
     \hspace*{-0.9in}
    \includegraphics [scale=0.65] {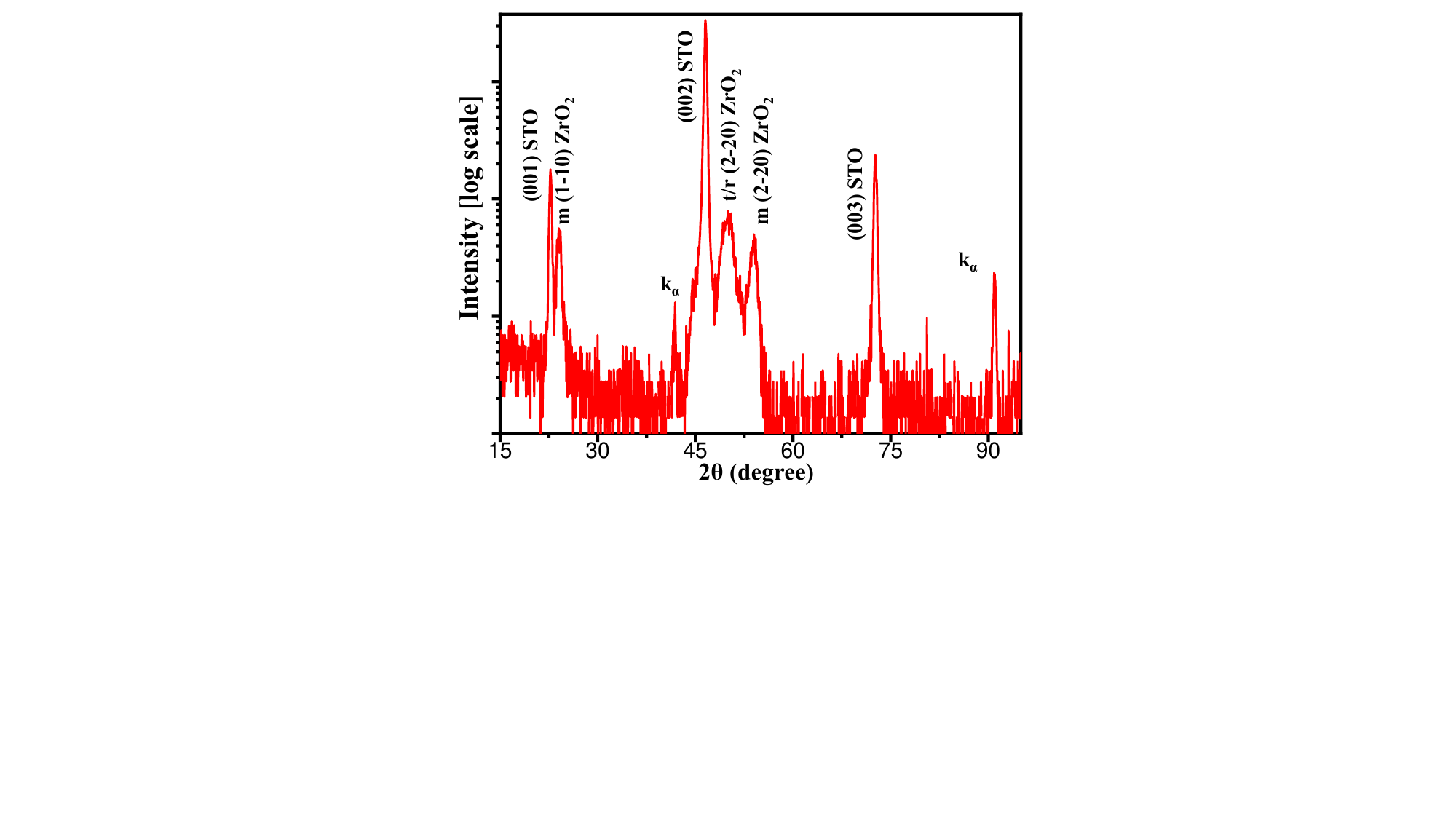}
    \vspace*{-53mm}
    \caption{In-plane X-Ray diffraction, $\theta - 2 \theta$ scan of 42 nm-thick ZrO$_2$ thin film grown on LSMO-buffered 110-oriented STO substrate.}
    \label{fig:S4} 
\end{figure}

Figure \ref{fig:S2} shows the $\theta - 2\theta$ Bragg scans for all the spots observed on the pole figure shown on the left. The $\theta - 2\theta$ values are fitted with a double Gaussian to extract the lattice parameter values, and the results are displayed in Figure \ref{fig:S3}. The averages of the 12 extracted values are also shown, giving a$\sim$b = 5.097 Å and c = 5.193 Å.

\noindent \textbf{42 nm thick-ZrO$_{2}$.}\\
Figure \ref{fig:S4}  presents the in-plane XRD measurement along the <001> azimuthal direction (<001>-STO). It can be seen that the monoclinic phase is present and can be easily detected (as observed for pure HfO$_2$ as shown below). The in-plane measurement shows greater sensitivity than the out-of-plane one in detecting phases that are present in small amounts. In the main paper (Figure 1), it can be observed that the intensity of the monoclinic phase in 42 nm-thick ZrO$_2$ thin film is low, while it is significant in the in-plane measurement.

\begin{figure}[H]
    \centering
     \hspace*{-0.9in}
    \includegraphics [scale=0.6] {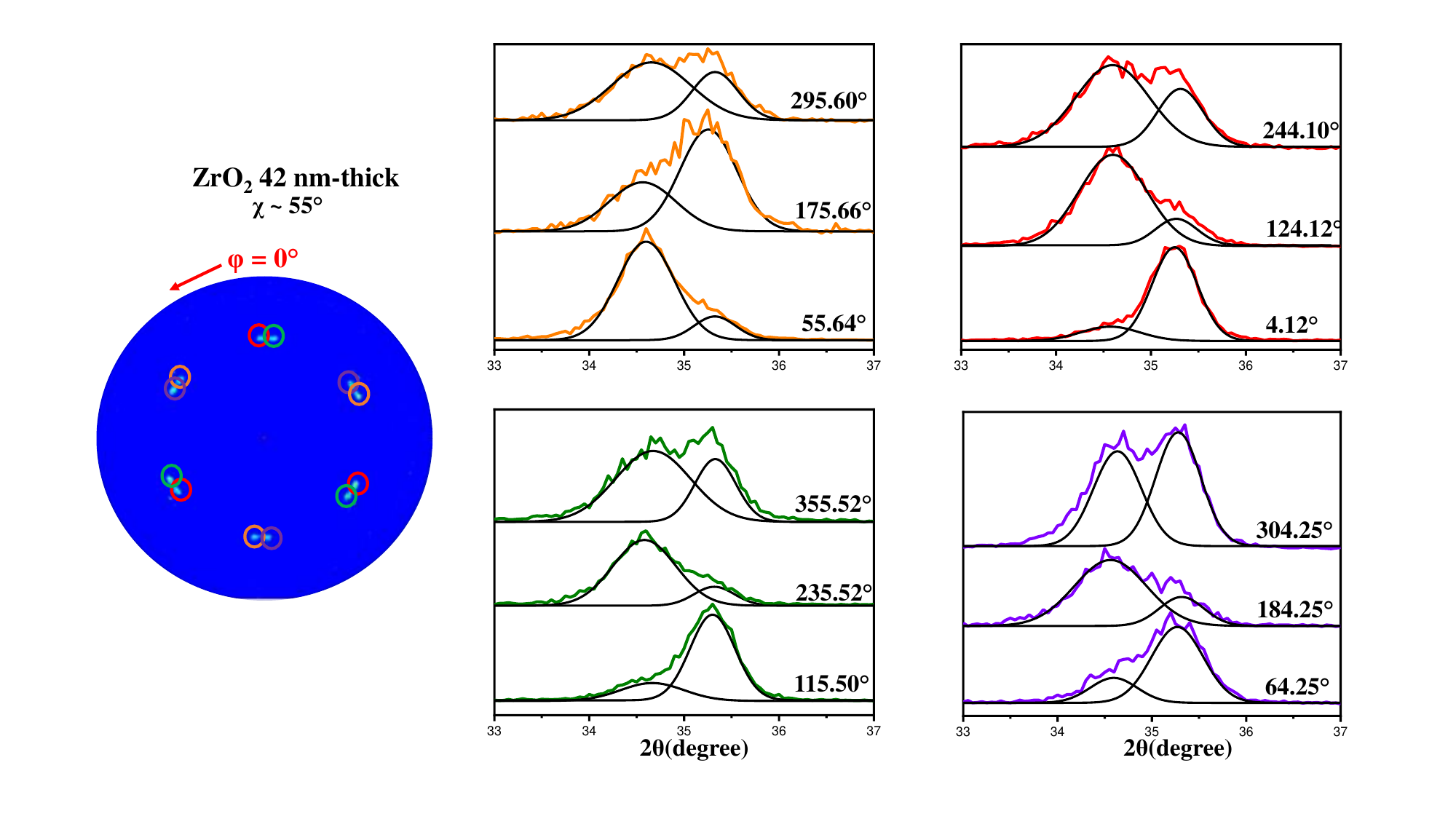}
    \vspace*{-8mm}
    \caption{X-ray diffraction pole figure generated at $2 \theta =$ 34.5°. The Bragg $\theta - 2 \theta$ scan for each spot observed in the pole figure is shown on the right.}
    \label{fig:S5} 
\end{figure}

\begin{figure}[H]
    \centering
     \hspace*{-0.5in}
    \includegraphics [scale=0.45] {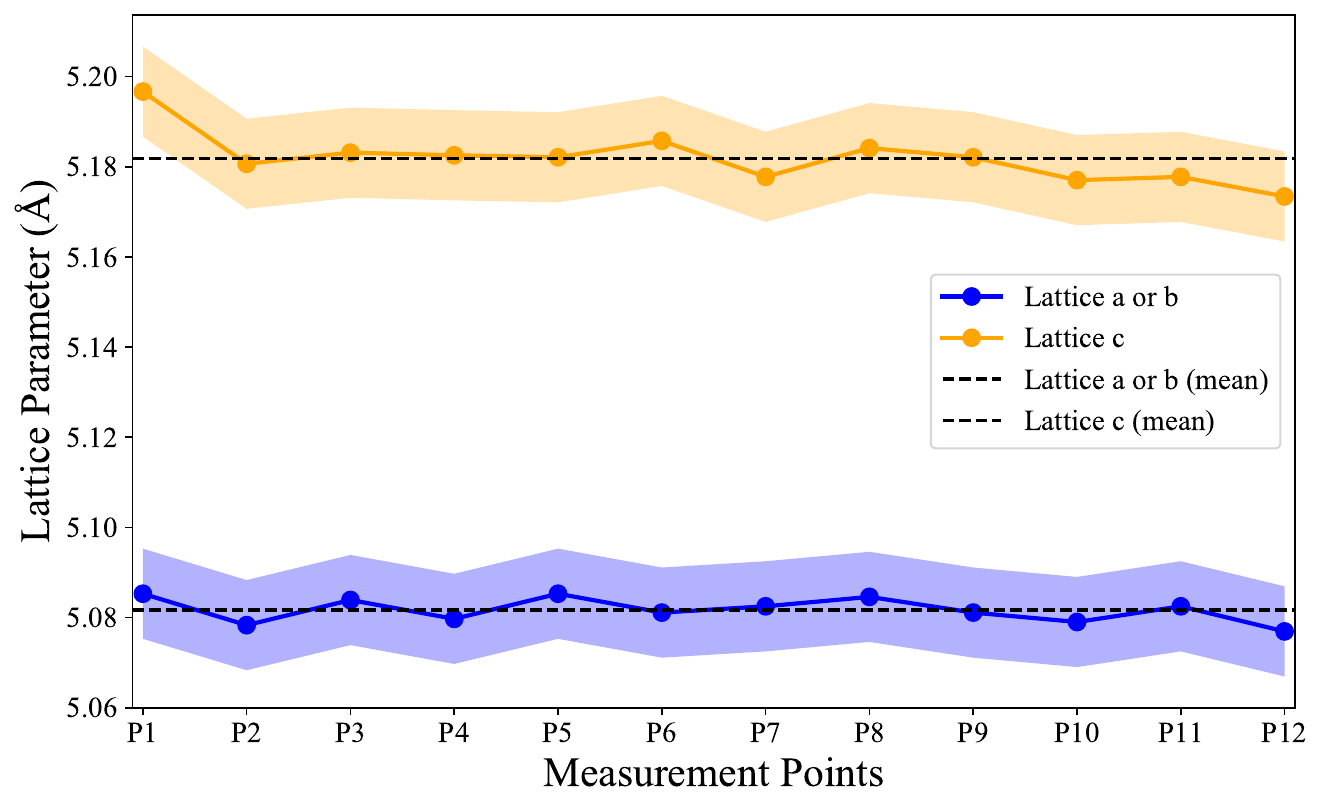}
    \vspace*{-2mm}
    \caption{The extracted a, b, and c parameters from Figure \ref{fig:S5}, with P1 to P12 corresponding to spots at $\phi =$ 4.12° to $\phi =$ 355.52°, respectively.}
    \label{fig:S6} 
\end{figure}

\begin{figure}[H]
    \centering
     \hspace*{-0.6in}
    \includegraphics [scale=0.55] {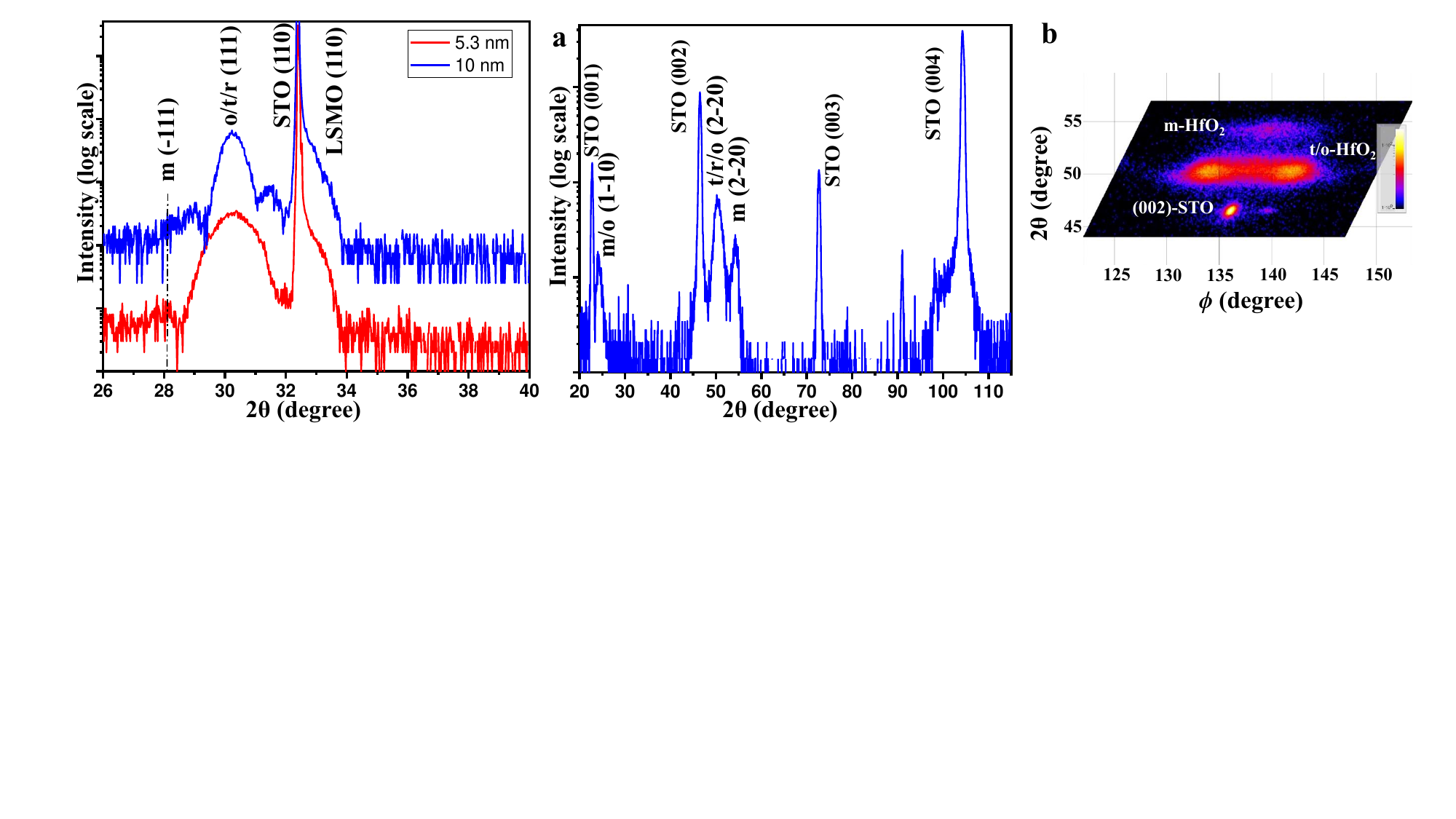}
    \vspace*{-54mm}
    \caption{Out-of-plane and in-plane X-Ray diffraction for pure HfO$_2$: (a) Out-of-plane $\theta - 2 \theta$, (b) in-plane $\theta - 2 \theta$/$\chi$, (c) 2D map ($\phi-$ 2 $\theta$/$\chi$) around (001)-STO plane. Thickness is 10 nm (in blue and for 2D map) and 5.3 nm (in red).}
    \label{fig:S7} 
\end{figure}

\begin{figure}[H]
    \centering
     \hspace*{-0.15in}
    \includegraphics [scale=0.55] {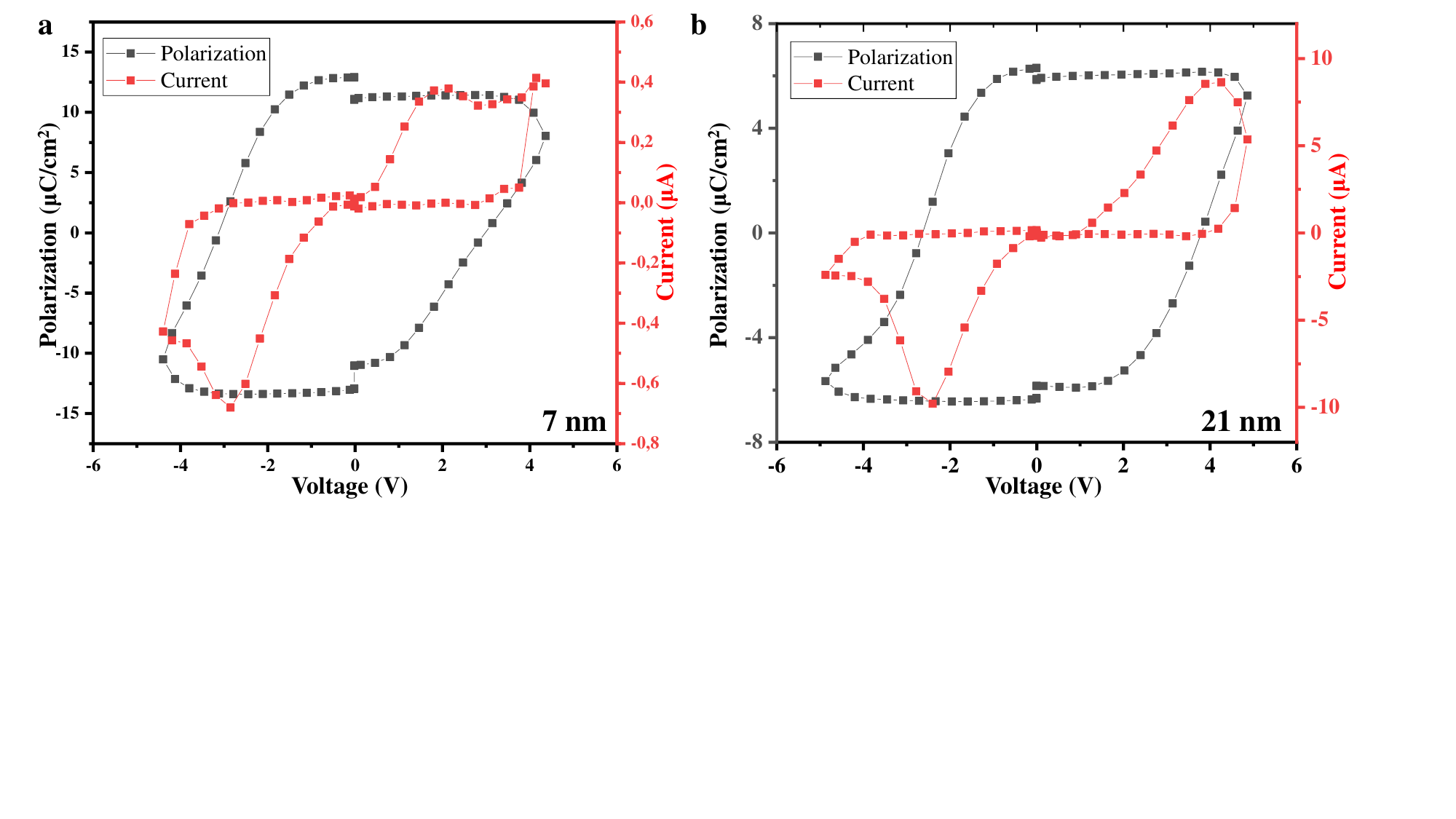}
    \vspace*{-44mm}
    \caption{P-E Polarization hysteresis loops of 7 and 21 nm-thick pure ZrO$_2$ thin films grown on LSMO-buffered 110-oriented STO substrate.}
    \label{fig:S8} 
\end{figure}

\begin{figure}[H]
    \centering
     \hspace*{-0.15in}
    \includegraphics [scale=0.55] {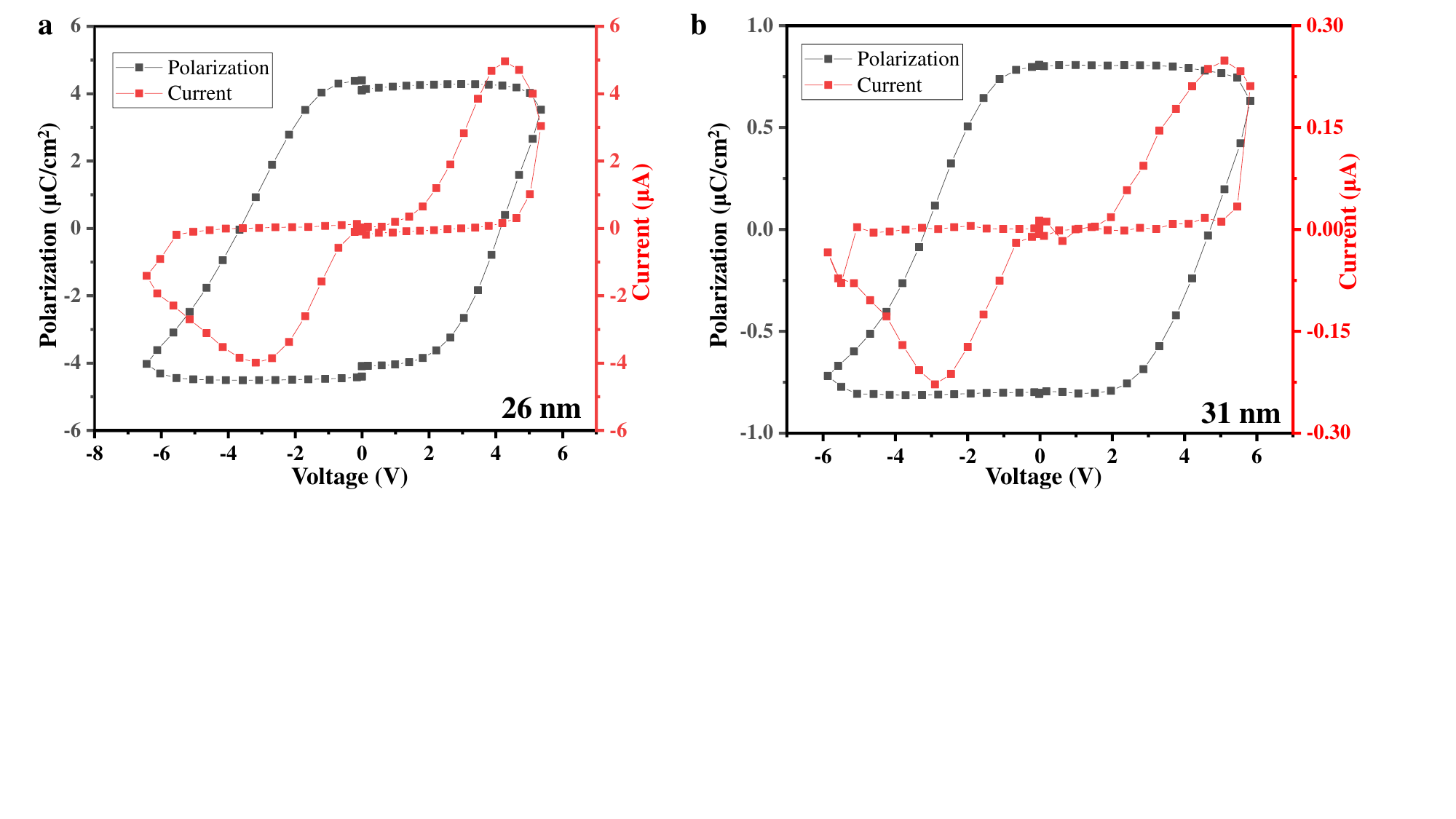}
    \vspace*{-44mm}
    \caption{P-E Polarization hysteresis loops of 26 and 31 nm-thick pure ZrO$_2$ thin films grown on LSMO-buffered 110-oriented STO substrate.}
    \label{fig:S9} 
\end{figure}

Figure \ref{fig:S5} shows the $\theta - 2\theta$ Bragg scans for all spots observed in the pole figure (also given in the same figure). The $\theta - 2\theta$ values are fitted with a double Gaussian to extract the lattice parameter values, and the results are displayed in Figure \ref{fig:S6}. The averages of the 12 extracted values are also shown, giving a$\sim$b = 5.082 Å and c = 5.183 Å.
\noindent \textbf{Pure HfO$_{2}$.}\\
Figure \ref{fig:S7}.a presents the out-of-plane $\theta-2\theta$ scans of 5.3 and 10 nm-thick pure HfO$_2$ thin films. As shown in Figure \ref{fig:S7}.a, it is not easy to detect the presence of a monoclinic phase, although in pure HfO$_2$ the most stable phase is monoclinic. However, in the in-plane XRD measurement (Figure \ref{fig:S7}.b), the presence of a monoclinic phase is clearly detected, similar to what was observed in 42 nm-thick ZrO$_2$ films (see above). Figure \ref{fig:S7}.c shows a $\phi-2\theta/\chi$ 2D map, which reveals that the growth of HfO$_2$ follows the same mechanism as pure ZrO$_2$ (as reported in the main paper for pure ZrO$_2$). The only difference seems to be the thickness stability of the t phase, which is higher in pure ZrO$_2$ than in HfO$_2$.
\noindent \textbf{Electric measurements.}\\
For the P-E loop measurements, a simple Platinum/film/LSMO/STO capacitor was used. The P-E loops were measured using the PUND method. Figures \ref{fig:S8} and \ref{fig:S9} display the electric measurement results for 7, 21, 26, and 31 nm-thick ZrO$_2$ thin films, which all exhibit clear ferroelectric behavior. The thinnest film of 7 nm presents much more leakage current than the thicker ones.

\end{document}